\PassOptionsToPackage{table}{xcolor}
\documentclass[sigconf]{acmart}

\usepackage{listings}
\usepackage{enumitem}
\usepackage{stfloats}
\usepackage{tcolorbox}
\usepackage{soul}

\definecolor{placeholdergray}{RGB}{150,150,150} % Light gray

\newcolumntype{Y}{>{\raggedright\arraybackslash}X}
\AtBeginDocument{%
  }

\copyrightyear{2026}
\acmYear{2026}
\setcopyright{cc}
\setcctype{by}
\acmConference[SIGCSE Virtual 2026]{Proceedings of the 2nd ACM Virtual Global Computing Education Conference V.1}{November 12--15, 2026}{Virtual Event, USA}
\acmBooktitle{Proceedings of the 2nd ACM Virtual Global Computing Education Conference V.1 (SIGCSE Virtual 2026), November 12--15, 2026, Virtual Event, USA}
\acmDOI{10.1145/3795867.3831014}
\acmISBN{979-8-4007-2506-7/2026/11}

\begin{document}

\title{LLM-Generated Design Problems for Assessing Higher-Order Thinking in
Project-Based Learning}

\author{Ahmad D. Suleiman}
\email{as4300@rit.edu}
\orcid{0009-0006-5144-1713}
\affiliation{%
  \institution{Rochester Institute of Technology}
  % \department{Computing and Information Sciences}
  \city{Rochester}
  \state{NY}
  \country{USA}
}
\author{Daqing Hou}
\email{dqvse@rit.edu}
\orcid{0000-0001-8401-7157}
\affiliation{%
  \institution{Rochester Institute of Technology}
  % \department{Department of Software Engineering}
  \city{Rochester}
  \state{NY}
  \country{USA}
}
\author{Maliha Noushin Raida}
\email{mr6074@rit.edu}
\orcid{0000-0002-1729-2930}
\affiliation{%
  \institution{Rochester Institute of Technology}
  % \department{Computing and Information Sciences}
  \city{Rochester}
  \state{NY}
  \country{USA}
}

\begin{abstract}
\label{section:abstract}

Project-based learning (PjBL) is common in computing education, but traditional assessments of PjBL often fail to capture higher-order thinking (HOT), especially in transfer contexts. This study introduces ``design problems'' (DPs): concise, scenario-based prompts that require applying project concepts in new situations, to address this gap. We examined instructor perceptions, the ability of large language models (LLMs) to generate DPs, and student experiences. Surveys of 31 instructors, evaluation of 80 LLM-generated DPs, and student performance data showed that while instructors value DPs, creation effort is a barrier. LLMs helped by producing high-quality prompts with strong expert agreement.
Students rated DPs from different LLMs similarly, and their performance on DP tasks showed negligible correlation with traditional project grades, suggesting DPs may capture distinct aspects of HOT. Keystroke data also suggested deeper cognitive engagement of students through planning and revision behaviors. Overall, DPs appear to be a useful complement to traditional assessments, especially in situations where AI use or collaboration may undermine individual learning.
\end{abstract}




\maketitle

\section{Introduction}
\label{section:introduction}

% GOOD DP Example
\begin{figure}[t]
    \centering
        \begingroup
    \setlength{\fboxsep}{2pt}
    \setlength{\fboxrule}{0.5pt}
    \fbox{%
    \begin{minipage}{\dimexpr\linewidth-2\fboxsep-2\fboxrule\relax}
    \small
    \textbf{Scenario}:
    
    \sethlcolor{orange!30} \hl{A healthcare technology company wants to create an application to help families and caregivers monitor the well-being of elderly individuals. One early indicator of certain health issues is a change in fine motor skills, which refers to the small, precise movements of the hands and fingers.} The company believes that how a person interacts with their smartphone's touchscreen could reveal subtle, long-term changes in these skills.
    \medskip

    \textbf{Problem}:   
    
    Propose a \sethlcolor{red!30} \hl{high-level design} for a system that uses a person's daily smartphone usage to detect potential declines in their motor skills. Your description should outline the \hl{major components of your system} and explain \sethlcolor{cyan!40} \hl{how information would flow between them} to achieve the goal. Specifically, describe (1) \hl{how you would capture relevant user interaction data from the phone}, (2) \sethlcolor{green!30} \hl{how this information would be stored and managed centrally}, and (3) \sethlcolor{cyan!40} \hl{how you would process this information to establish a personal baseline and identify significant, long-term deviations from it}.

    \end{minipage}
    }
    \endgroup
    \vspace{-1em}    
    \caption{\small Example of high-quality LLM-generated DP from  Touchalytics project. \textcolor{orange!70!white}{Orange} highlights show a request to \textit{transfer} from the Touchalytics concepts to a new, socially relevant elder-care context. The problem provides opportunities to elicit HOT skills (\textit{analyzing} interaction signals \textcolor{cyan!90!white}{Blue}, \textit{evaluating} privacy and storage tradeoffs \textcolor{green!60!black}{Green}, and \textit{synthesizing} an architecture for the system \textcolor{red!100!black}{Red}).}
    \vspace{-5mm}
    \Description{A high-quality LLM-generated design problem showing transfer and HOT through analyzing, evaluating, and synthesizing an end-to-end system.}
    \label{fig:good_problem}

\end{figure}

Project-based Learning (PjBL) is widely adopted in computing education due to its effectiveness in fostering practical skills and deeper cognitive engagement~\cite{ahmadSLR}. Traditional PjBL assessment practices are mainly based on project artifacts, such as code, reports, and presentations~\cite{tenhunen2023systematic}. However, significant challenges remain in accurately assessing students' higher-order thinking (HOT) skills in \textit{transfer contexts beyond the original project} (i.e., analysis, evaluation, and creation in Bloom's taxonomy~\cite{bloom}). Although some educators supplement PjBL assessment with final exams, these often target lower-order cognitive skills and/or lack alignment with the original project context~\cite{imbulpitiya2021examining, jacobson2024right}. Continuous assessment approaches that track individual progress are potentially valuable but can be impractical due to high demands on instructor time and resources, especially in large classes~\cite{garcia2018pilot}.

In this study, we evaluate the potential of using what we call \textit{design problems (DPs)} as a complementary assessment approach to address these limitations. A well-crafted DP aligns with the original project learning objectives (LOs) but is presented in a concise, realistic, distinct scenario. Fig.~\ref{fig:good_problem} illustrates an  LLM-generated DP. Drawing on transfer of learning theory~\cite{perkins1992transfer}, our main purpose is to assess whether students can transfer knowledge and skills learned from PjBL to new contexts. This approach goes beyond simple recall or understanding, instead designed to require students to engage in HOT skills. Furthermore, DPs are meant to be completed in a short period of time, so they can be used in various settings, such as exams, interviews, or presentations. To keep grading manageable while still ensuring meaningful responses, student answers are expected to be a few sentences or a short paragraph. 

In addition, DPs can address two growing challenges in assessing PjBL. First, the widespread use of generative AI tools, such as ChatGPT~\cite{prather2024widening}, allows students to superficially complete project tasks without truly mastering critical cognitive skills. Second, the collaborative nature of PjBL complicates attribution of individual contributions and assessment of individual mastery~\cite{hundhausen2022assessing}. Because DPs require individual responses beyond simple recall or understanding, they can more effectively evaluate each student's learning and HOT skills, especially in environments where AI use and teamwork often obscure true mastery.

However, creating high-quality DPs tailored to specific student projects and LOs can be labor-intensive and demand considerable instructor effort. This bottleneck limits the scalable, consistent creation of DPs in computing education. Fortunately, recent advances in large language models (LLMs) offer a possible solution, as research has highlighted their ability to generate relevant educational materials that reduce instructor workload while maintaining high pedagogical quality~\cite{raihan2025large, logacheva2024evaluating}. Specifically, LLMs have shown the potential to generate contextualized programming exercises that effectively target HOT skills~\cite{del2024evaluating, al2024analysis}.

Given these challenges and opportunities, this study investigates the following key research questions (RQs): 

\begin{enumerate}
  \renewcommand{\labelenumi}{\textbf{RQ\arabic{enumi}:}}
    \item \textit{How do computing instructors perceive the value and practicality of using DPs to assess HOT in PjBL?} 
    
    \item \textit{Can LLMs effectively generate high-quality DPs for PjBL transfer-oriented HOT assessment?}
    
    \item \textit{How do students experience LLM-generated DPs when deployed in authentic PjBL classrooms?}
\end{enumerate}
\section{Related Work}
\label{section:related_work}

\subsection{Assessment of Transfer of Learning}
% DPs as Assessment Tools
\label{subsection:design_problems}
Our approach is grounded in transfer of learning theory~\cite{perkins1992transfer} to address the need to assess HOT in PjBL~\cite{grossman2022core}. Transfer occurs when knowledge acquired in one context is applied in a new one. In PjBL, students develop deep knowledge through extended engagement with a specific project, but assessments of true mastery require the ability to recognize when and how to apply that knowledge beyond the original learning context~\cite{Barron01071998}. DPs assess this transfer capability by presenting scenarios that share underlying principles with the original project but differ in surface features, domain, or constraints. Success requires abstracting core concepts from the original context and applying them to new requirements. This far transfer represents a deeper understanding than simply reproducing solutions, directly targeting the upper levels of Bloom's taxonomy~\cite{bloom}. This framing motivates DPs as a form of transfer-oriented assessment, but raises a practical challenge: creating high-quality, project-specific DPs requires substantial instructor effort.

\subsection{LLMs in Computing Education}
The advancement of LLMs has driven extensive research into their potential applications within computing education. Early explorations leveraged LLMs to support code-related educational tasks, such as generating code, debugging, code reviews, and grading~\cite{prather2023robots}. Recent work~\cite{raihan2025large,dangol2026relief} illustrates broader educational applications, noting LLMs' usefulness in enhancing learning experiences, supporting curriculum development, and generating educational content. Automatic question generation has emerged as a significant area of interest. Traditional techniques relied heavily on template-based approaches or neural networks trained on predefined question sets, often resulting in generic or contextually limited outcomes~\cite{serban2016generating, duan2017question}. In contrast, LLMs offer more context-aware and flexible question generation capabilities. Tran et al.~\cite{tran2023generating} demonstrated ChatGPT's proficiency in generating high-quality multiple-choice questions aligned with a programming course. Beyond multiple-choice questions, LLMs have proven effective in generating richer, contextualized exercises, which consist of a narrative background and a programming task along with solution code that educators and students perceived as clear, relevant, and beneficial~\cite{del2024evaluating,logacheva2024evaluating}.

Our work builds on previous research by focusing on the generation of DPs tailored to the LOs of a given PjBL project. Recognizing that the development of effective HOT assessments requires substantial efforts~\cite{hubbard2017question}, we aim to leverage LLMs to reduce the effort in creating high-quality DPs.

\section{Research Methods}
\label{section:methods}

\subsection{Instructor Survey (RQ1)}
To address RQ1, we conducted a brief, mostly open-ended survey.
\subsubsection{Recruitment}
We targeted instructors with experience in PjBL by emailing 313 authors of papers in a recent computing PjBL literature review~\cite{ahmadSLR}. Publishing on computing PjBL suggests practical experience in designing and assessing PjBL courses. The Google Forms survey stayed open for one month, and we sent two reminder emails. Participation was voluntary and anonymous. Respondents provided informed consent under an IRB-approved protocol. The survey took about five minutes to complete.

% Table for Instructor Questionnaire
\begin{table}[t]
    \centering
    \caption{Questionnaire: Instructors' Perceptions of DPs}
    \label{tab:design-problem-questionnaire}
    \small
    \rowcolors{2}{gray!10}{white}
    \begin{tabular}{@{}p{0.74\columnwidth}p{0.2\columnwidth}@{}}
    \toprule
    \textbf{Question} & \textbf{Response}\\
    \midrule
    Consent Form & Yes/No\\
    \hline
    
    Preamble: Description of HOT; Example of Design Problem with project context (description and objectives) & N/A\\
    
    Based on the explanation and example above, are our notions of HOT and design problems clear to you? & Yes/No/Other\\
    \hline
    
    How do you assess student learning outcomes in your project-based computing course(s)? & Open-ended\\
    
    Do you explicitly assess HOT skills (e.g., analysis, evaluation, creation)? How? & Open-ended\\
    
    \hline
    Do you believe design problems can be effective for assessing HOT skills in a project-based computing course? & Yes/No/Other\\
    
    Please leave comments, questions, and concerns, if any: & Open-ended\\
    
    \bottomrule
    \end{tabular}
    % \vspace{-4mm}
\end{table}

\subsubsection{Instrument design}
To collect instructor perceptions of DP, we developed a brief questionnaire (Table~\ref{tab:design-problem-questionnaire}). To support a common interpretation, the questionnaire first included a preamble defining HOT and introducing DPs with an example. It then gathered background information about instructors’ current PjBL assessment practices and whether they explicitly assess HOT skills. Finally, it collected instructors' perspectives, concerns, and anticipated value regarding DPs as an assessment tool.

\subsubsection{Analysis}
In total, 31 instructors completed the survey (\textasciitilde10\% response rate). Two co-authors analyzed responses using reflexive thematic analysis~\cite{braun2006using}, appropriate for characterizing instructors’ perceptions and concerns. They first read all responses to become familiar with the data, then coded responses inductively. After the initial coding pass, the two coders compared interpretations, discussed overlapping or ambiguous codes, and collaboratively revised the code set. See Section~\ref{result:survey} for the survey results.

\subsection{LLM DP Generation and Evaluation (RQ2)}
\label{subsection:llm_generation}

\subsubsection{Project Selection}
To generate DPs using LLMs, we selected four existing PjBL projects that have been taught in multiple undergraduate computing courses (Table~\ref{tab:projects}). We chose these projects to maximize diversity in \textit{domain} and \textit{scale}: web security, distributed systems, software maintenance, and mobile machine learning, with varying complexity and cognitive demands.

% Table for List of projects and their corresponding deployment details
\begin{table}[t]
\caption{Projects  for LLM-based DP Generation and Deployment (Courses and DP exam Web Link  redacted for review)}
\label{tab:projects}
\small
\centering
\rowcolors{2}{gray!10}{white}
\begin{tabular}{@{}p{1.3cm}p{\dimexpr\columnwidth-1.3cm-2\tabcolsep\relax}@{}}
\toprule
\textbf{Project} & \textbf{Details} \\
\midrule

Breached! & 
Web service to verify if user credentials appear in breached datasets, using cryptographic hashing.
\textit{Primary LO}: Design solution for securing personal data by applying concept of hashing. 
\textit{DP Exam:} Course~A, 3rd \& 4th Year,
$N=28$, Individual Project,
In class, WebApp-based exam \\

Distributed Tic-Tac-Toe (DT3) & 
Two-player distributed game with synchronized gameplay and persistent data management.
\textit{Primary LO}: Design client-server application with socket programming and multithreading.
\textit{DP Exam:} Course~B, 4th Year \& Grad.,
$N=40$, Team of 5-6,
In class, Paper-based exam \\

Moodle & 
A maintenance project for the Moodle LMS involving new feature development, enhanced security, and usability.
\textit{Primary LO}: Analyze, design, and improve software systems by applying software change process.
\textit{DP Exam:} N/A \\

Touchalytics & 
Mobile app that uses swiping behavior to authenticate users.
\textit{Primary LO}: Design solutions that synthesize concepts from mobile development, behavioral biometrics, and machine learning. 
\textit{DP Exam:} Course~C, 4th Year,
$N=10$, Team of 5-6,
In class, WebApp-based exam  \\
\bottomrule
\end{tabular}
\vspace{-1.5em}

\end{table}

\subsubsection{Prompt Engineering}
We employed a zero-shot prompt engineering~\cite{phoenix2024prompt} approach to preserve diversity and avoid exemplar bias. Prior work~\cite{del2024evaluating} on contextualized programming exercise generation shows that this approach can produce pedagogically appropriate exercises without examples. In addition to project details, we included a target LO as an additional dynamic input to maintain focus. The final prompt (Fig.~\ref{fig:llm-prompt-template}) guided the model to generate DPs that require short student answers, explain domain-specific or unfamiliar terms, and avoid any hints at a solution. To ensure consistency and statelessness, all generations were executed via the API rather than a user interface. To balance diversity and hallucination, we set the temperature parameter to 0.7~\cite{chen2025unleashing}.
 
% LLM Prompt Figure
\begin{figure}[t]
\centering
\begingroup
\setlength{\fboxsep}{2pt}
\setlength{\fboxrule}{0.5pt}
\fbox{%
\begin{minipage}{\dimexpr\linewidth-2\fboxsep-2\fboxrule\relax}
\footnotesize
You are an expert computing education assistant. 
You will generate a short exam design problem for students to solve.
Format Requirements:
\begin{itemize} [leftmargin=*]
    \item The problem should begin with a short, realistic scenario.
    \item The problem should test HOT skills
    \item The problem should be in the scope of the project, but in a different scenario.
    \item The expected answer length is a few sentences to one paragraph.
    \item Students should be able to answer in 30 minutes.
    \item Do not provide any technical terms that could serve as hints to the solution.
    \item Ensure to adequately explain concepts in the domain problem that are potentially unfamiliar to students.
\end{itemize}

Generate a distinct problem that aligns with these requirements based on the following project details:
\begin{enumerate} [leftmargin=*]
    \item Project Overview \& Learning Objectives: 
    \textcolor{placeholdergray}{\texttt{<learning-objectives>}}
    
    \item Project and Tasks Description: 
    \textcolor{placeholdergray}{\texttt{<project-description>}}
    
    \item Key Technologies, Techniques, and Tools for the Project: 
    \textcolor{placeholdergray}{\texttt{<technologies>}} 
    
    \item Target Learning Objectives of the Design Problem: 
    \textcolor{placeholdergray}{\texttt{<target-objectives>}}

\end{enumerate}
Output format:

Scenario: $<$Brief context description$>$

Problem: $<$Specific problem posed to students$>$

\end{minipage}
}
\endgroup
\vspace{-1em}
\caption{LLM Prompt Template for Generating DPs.}
\Description{LLM Prompt Template for Generating Design Problems to Assess HOT in Computing PjBL.}
\vspace{-2mm}
\label{fig:llm-prompt-template}
\end{figure}

\subsubsection{Generation Process}
For each project, we generated DPs using two LLMs: OpenAI GPT-4o and a reasoning model, Gemini 2.5 Pro (as OpenAI o3 was unavailable to us via API). We included a reasoning-oriented LLM (sometimes referred to as ``thinking'' models) to explore its capability to generate DPs. Our aim is not to compare the two brands, but to compare a standard LLM with a reasoning-oriented one. For each project, we generated 10 DPs with each LLM, resulting in 20 DPs per project and 80 DPs overall.

\subsubsection{Expert Evaluation}

% Table for Expert Evaluation Rubric
\begin{table}[t]
\caption{Evaluation Criteria for  LLM-Generated DPs}
\label{tab:llm-eval-rubric}
\small
\centering
\rowcolors{2}{gray!10}{white}
\begin{tabular}{@{}p{1.7cm}p{\dimexpr\columnwidth-1.7cm-2\tabcolsep\relax}@{}}
\toprule
\textbf{Criterion} & \textbf{Description}\\
\midrule

\textbf{SQ}: Scenario Quality&
If the DP scenario is realistic, contextually rich, and useful for helping students transfer project knowledge.\\

\textbf{AL}: Alignment&
How closely the DP reflects the learning objectives, without omitting any or introducing unrelated skills.\\

\textbf{CC}: Cognitive Complexity&
If the DP engages students in HOT (analyzing, evaluating, and designing) rather than only recalling facts.\\

\textbf{CS}: Clarity \& Specificity&
If the DP is clearly stated, self-contained, and free of ambiguities that could confuse or mislead students.\\

\textbf{FS}: Feasibility&
If the DP is appropriately scoped and solvable within 30 min by students with expected prior knowledge.\\

\bottomrule
\end{tabular}
\vspace{-1.5em}
\end{table}
We developed a rubric (summarized in Table~\ref{tab:llm-eval-rubric}) informed by prior studies evaluating LLM-generated exercises~\cite{del2024evaluating, logacheva2024evaluating}. We adopted and refined five criteria to suit our study: scenario quality, alignment with LOs, cognitive complexity, clarity/specificity, and feasibility. The wording of each criterion was adapted to reflect the DP context while maintaining conceptual consistency with prior work. The research team iteratively refined the rubric to ensure that each criterion captured a distinct aspect of DP quality. Before formal evaluation, two expert raters (authors of this work) piloted the rubric on 10 generated DPs and clarified the scope of each criterion. Each criterion was rated as Yes (1), Maybe (0.5), or No (0), with brief rationale comments. The two raters then independently evaluated all 80 DPs using the finalized rubric.

\subsubsection{Analysis}
Cohen’s $\kappa$ with quadratic weights~\cite{cohen1968weighted} was used to assess inter-rater reliability (as the scale is ordinal). To perform statistical comparisons, we averaged the two ratings for each criterion. We then used independent two-tailed $t$-tests~\cite{welch1947generalization} and Cohen’s $d$~\cite{cohen1988statistical} effect sizes to compare DP quality between the two LLMs, and one-way ANOVA~\cite{girden1992anova} with Tukey’s HSD post hoc tests~\cite{tukey_hsd} to compare DP quality across projects. See Section~\ref{result:llm} for the qualitative review and model/project analysis.

\subsection{Classroom Deployment (RQ3)}

To investigate how students experience LLM-generated DPs in an authentic PjBL setting (RQ3), we deployed DPs in three undergraduate computing courses (see Table~\ref{tab:projects}). All deployments were 30-minute supervised in-class summative assessments to reduce unauthorized assistance and ensure individual work. Two courses (Course~A - Fundamentals of Software Engineering, using \textit{Breached!} and Course~C - Senior Software Engineering, using \textit{Touchalytics}) used a custom web-based assessment application to deliver the DPs in an exam-style format, capture written responses, and log timestamped keystrokes. The third course (Course~B - Computer Networks, using \textit{DT3}) used paper format due to logistical constraints.

\subsubsection{Assessment Design}
Each student completed two tasks. \textit{Task~A} required students to answer one DP randomly selected from a pool of high-quality DPs with strong expert ratings. \textit{Task~B} required students to rate another DP randomly selected from the full generated bank (including both high- and low-rated DPs). This design enabled us to collect both performance data and student perceptions of the DP quality. Students rated the DP using the same five criteria used by expert raters (Table~\ref{tab:llm-eval-rubric}), with wording adapted for students and presented on a 5-point Likert scale ranging from ``Strongly Disagree'' (1) to ``Strongly Agree'' (5). To incentivize genuine effort, this assessment contributed 2\% to students' final course grades.

\subsubsection{DP Performance Grading}
Student responses to Task~A were graded by the authors rather than course instructors using an 8-point rubric for consistency across courses. The \textit{Solution} criterion (0-6 points) evaluated higher-order Bloom’s levels, including \textit{creating} a technically sound design proposal, \textit{analyzing} scenario constraints to identify relevant requirements and limitations, and \textit{evaluating} alternatives through justified reasoning that linked design decisions to contextual factors. The \textit{Clarity} criterion (0-2 points) assessed response organization and readability.

\subsubsection{Data Collection}
We collected three types of data. First, \textit{performance data} from Task~A grades captured students' DP performance. Second, \textit{perception data} from Task~B ratings, showing how students evaluated DP quality across the five criteria. Third, for the two web-based deployments, we collected \textit{behavioral data} through keystroke logging, capturing timestamped character additions and deletions. Keystroke logging is well-established for collecting time-based traces of writing behavior~\cite{TIAN2025100179}, enabling analysis of cognitive engagement patterns such as planning, revision, and thinking pauses. In the web-based deployments, we also reviewed keystroke traces as an integrity check to confirm normal typing and revision patterns rather than abrupt copy-paste insertions.

\subsubsection{Analysis}
For \textit{performance data}, we compared Task~A grades across the three courses using one-way ANOVA. We calculated Spearman's correlations between Task~A grades and word count, time on task, and perception ratings. For \textit{perception data}, we summarized Task~B ratings using descriptive statistics and tested model and course differences using $t$-tests and ANOVA. We also compared student and expert ratings on the 39 shared DPs using weighted Cohen’s $\kappa$ and paired $t$-tests. For \textit{behavioral data}, we extracted three keystroke metrics: (1) \textit{initial planning latency}, time from task start to first keystroke; (2) \textit{revision ratio}, characters deleted divided by characters added; and (3) \textit{burstiness}, frequency of pauses exceeding 10 seconds. These metrics were interpreted as indirect behavioral proxies for planning, revision, and sustained thinking during DP solving, drawing on writing-process theory and pause-analysis research~\cite{flower1981cognitive,schilperoord2002pauses,TIAN2025100179}. See Section~\ref{result:deployment} for the deployment results.

\subsubsection{Relationship to Traditional Project Grades}
To examine how DP performance relates to traditional project grades, we compared Task~A grades with project grades using Spearman correlation and learner profiling. For learner profiling, students were categorized into four profiles based on a performance matrix, defining ``high project performance'' as scoring $\ge$ 95\% on the project grade and ``high design performance'' as scoring $>50\%$ (a passing grade) on the DP, yielding \textit{Masters} (high project/high design), \textit{Conceptualizers} (low/high), \textit{Implementers} (high/low), and \textit{Strugglers} (low/low).
\vspace{-1mm}
\section{Results}
\label{section:result}

\subsection{RQ1: Instructors’ Perspectives}
\label{result:survey}
This section presents the results of the instructor survey ($N=31$).

\subsubsection{Current Assessment Practices}
To confirm that our sample reflects assessment practices reported in prior literature~\cite{tenhunen2023systematic}, we asked instructors which assessment forms they used. The responses aligned with the literature: evaluating artifacts (i.e., code and report) remains the dominant method (29 instructors; 93.5\%). Presentation-style assessments (e.g., demos or vivas) are also common (25; 80.7\%). In contrast, fewer instructors used traditional exams/quizzes (10; 32.3\%). Other methods included peer evaluation (12; 38.7\%) and progress tracking over time (5; 16.1\%).

\subsubsection{Explicit Targeting of HOT}
We asked instructors whether they explicitly assess HOT skills and how. Most (22 of 31; 71.0\%) said they do so explicitly. A small number (2; 6.5\%) described HOT as only implicitly assessed, and 7 (22.6\%) reported not targeting HOT, sometimes citing scale and workload constraints. However, even among those who explicitly assess HOT, few described practices that resemble the transfer-oriented approach we propose. Only 2 instructors (6.5\%) explicitly mentioned scenario-based questions that require students to analyze a new situation and propose solutions. Even with a broader interpretation to include slight variations of the original project, the transfer-oriented approach remained uncommon (4; 12.9\%). Instead, instructors more often relied on project artifacts, such as requiring design rationales and tradeoff justifications within reports (6; 19.4\%) or eliciting design reasoning through oral defenses, meetings, and demonstrations (11; 35.5\%). Overall, the results suggest a gap between instructors’ stated intent to assess HOT and the limited use of assessments that test transfer.

\subsubsection{Perceptions of DPs}
Three primary themes emerged regarding instructor perceptions of DPs. 
First, all 31 instructors supported the potential of DPs to assess HOT in PjBL. %No instructor opposed the concept. 
 Most indicated a clear agreement that DPs can be effective (22/31; 71.0\%), while the remainder agreed but with caveats (9/31; 29.0\%). 
Second, caveats centered on assessment validity, particularly with student use of generative AI. Several suggested that DPs are most convincing in settings requiring students to work independently, such as time-bound in-class assessments, and when paired with mechanisms that make student thinking visible, such as oral exams. This aligns with our intended use of DPs in exam-like settings in \textit{RQ3}.
Third, instructors raised adoption barriers related to quality and scalability. They emphasized that DPs must be carefully crafted to avoid simplistic prompts, support multiple plausible solutions, and match students' prior knowledge. They also noted the need for consistent grading and the burden of producing project-specific DPs across diverse projects. These concerns motivate \textit{RQ2}, where we use LLMs to reduce the creation bottleneck.

\subsection{RQ2: Evaluation of LLM-Generated DPs}
\label{result:llm}
Cohen's $\kappa$ across all criteria between the two expert raters yielded $\kappa=0.66$, indicating a substantial agreement~\cite{mchugh2012interrater}. Cohen's $\kappa$ for each criterion between the two raters also indicates a moderate to substantial agreement (SQ=.73, AL=.56, CC=.49, CS=.51, FS=.65).

\subsubsection{Qualitative Review}
LLMs generated realistic DP scenarios that go beyond what instructors might quickly draft, offering creative reframings of familiar projects into new domains while preserving core skills. For example, the DP shown in Fig.~\ref{fig:good_problem} requires students to \textit{transfer} concepts from the \textit{Touchalytics} project to a new context by repurposing the mobile behavioral data to elder-care health monitoring, giving students a fresh and socially relevant setting. The prompt explicitly requires students to engage in higher-order Bloom’s levels: \textit{analyze} the behavioral data to decide what to capture, \textit{evaluate} privacy and storage trade-offs, and \textit{create} an end-to-end architecture that detects long-term motor skills decline. We observed similar reframings across the projects, including \textit{Touchalytics} to keystroke-based identity verification, \textit{Moodle} to LMS features and usability improvements, \textit{DT3} to broader real-time client–server systems such as live polling, and \textit{Breached!} to security-driven domains such as protected radiology access and anonymous reporting. Overall, 40/80 generated DPs (50\%) received perfect scores from both raters across all five criteria.

On the other hand, some LLM-generated DPs fell short in several ways. A common issue (13/80) was scenarios being too similar to the original project, offering little meaningful difference from what students had already built. For example, in a \textit{Touchalytics} DP, the scenario restated the same mobile behavioral biometrics app students worked on during their project, which failed to create a genuine \textit{transfer} that requires adapting knowledge to a new context. Another issue was that some DPs (9/80) focused too narrowly on just one system aspect, limiting opportunities for students to critically think through design decisions and synthesize across systems. For instance, one \textit{Touchalytics} DP emphasized the machine learning component but did not require students to design how the mobile application supports that system, such as what swipe data to capture and how it is represented and stored. In some cases (4/80), prompts reduced cognitive demand by providing an obvious solution path or hints, limiting opportunities to \textit{analyze} alternatives, \textit{evaluate} tradeoffs, and justify design decisions. Clarity issues (16/80) further undermined these DPs when the expected scope was ambiguous, such as whether students should improve an existing system component or design a new one. Some DPs (6/80) were infeasible for a short assessment because they drifted into implementation-level details instead of design. When DPs demand overly technical details rather than architectural design choices, they become difficult to complete within the intended timeframe and increase grading workload. We believe many of these issues could be mitigated through targeted prompt engineering and minor changes by an instructor.

\subsubsection{Analysis by LLM and Project}

% Expert and Student Rating by Model for each criterion
\begin{table}[t]
 %   \vspace{-1.2em}
    \centering
    \small
    \caption{Expert mean ratings of standard and reasoning-oriented LLMs. ($d$ is Cohen's $d$. Bold* indicates $p<0.05$.)}
    
    \vspace{-0.5em}
    \label{tab:model_expert_student}
    \rowcolors{2}{gray!10}{white}
    \begin{tabular}{l | c c c c}
    \hline
    \textbf{Criterion} 
    & \textbf{Standard} & \textbf{Reasoning} & \textbf{$p$} & \textbf{$d$}\\
    \hline
    Scenario Quality        & 0.77 & 0.97 & \textbf{<0.001*} & \textbf{0.88}\\
    Alignment               & 0.96 & 0.94 & 0.548 & 0.14\\
    Cognitive Complexity   & 0.98 & 1.00 & 0.103 & 0.37 \\
    Clarity \& Specificity  & 0.93 & 0.94 & 0.851 & 0.04\\
    Feasibility             & 0.94 & 1.00 & \textbf{0.018*} & \textbf{0.55} \\
    \hline
    \end{tabular}
    % \caption*{\footnotesize \textbf{Note:} $p$ values are from two-tailed tests. \textbf{$d$} is Cohen's \textbf{$d$}. \textbf{Bold*} indicates $p<0.05$.}
    \vspace{-0.5em}
\end{table}

As shown in Table~\ref{tab:model_expert_student}, both the standard and reasoning-oriented LLMs produced high-quality DPs across criteria, suggesting that LLMs can generate prompts generally aligned with LOs and cognitively demanding. However, the reasoning-oriented LLM significantly outperformed the standard LLM on \textit{scenario quality} (SQ: 0.97 vs.\ 0.77, $p<0.001$, $d=0.88$), indicating more realistic and contextually rich scenarios supporting transfer. It also achieved higher \textit{feasibility} (FS: 1.00 vs.\ 0.94, $p=0.018$, $d=0.55$), suggesting a scope appropriate for a short assessment. No statistically significant differences were observed for alignment, cognitive complexity, or clarity/specificity ($p>0.05$).

Furthermore, expert ratings were consistently high across the four projects, suggesting generalizability across domains and project scales. Project-level averages across the five criteria ranged from 0.92 (\textit{Breached!}) to 0.97 (\textit{Moodle}), with \textit{DT3} (0.95) and \textit{Touchalytics} (0.94) in between. Alignment, cognitive complexity, and feasibility were uniformly strong across projects (AL: 0.93-1.00; CC: 0.96-1.00; FS: 0.96-0.99). Scenario quality remained strong overall (SQ: 0.80-0.93). The main project-level difference appeared in clarity, which varied significantly by project ($F(3,76)=4.88,\;p=.004$). Tukey HSD showed that both \textit{DT3} (CS=0.99) and \textit{Moodle} (CS=0.98) were significantly clearer than \textit{Breached!} (CS=0.84) ($p<0.01$).

\subsection{RQ3: Classroom Deployment}
\label{result:deployment}

% Table of Student Performnace + Time and Word Count
\begin{table}[t]

    \centering
    \caption{Student DP Performance in three Courses}
    \label{tab:student_performance}
    \small
    \rowcolors{2}{gray!10}{white}
    \begin{tabular}{l|ccccc}
        \hline
        \textbf{Course-Project} & \textbf{N} & \textbf{\(\bar{x}\) Grade} & \textbf{SD} & \textbf{\(\bar{x}\) Time} & \textbf{\(\bar{x}\) Words}\\
        \hline
        Course~A-Breached!       & 28 & 6.64 / 8 & 2.00 & 13.76m & 126.04 \\
        Course~B-DT3 & 40 & 3.93 / 8 & 2.84 & N/A   & 44.80 \\
        Course~C-Touchalytics    & 10 & 7.60 / 8 & 0.97 & 10.90m & 154.80 \\
        \hline
    \end{tabular}
        \vspace{-1.5em}

\end{table}

\subsubsection{Student Performance on DPs}

As shown in Table~\ref{tab:student_performance}, student DP performance differed significantly across the three deployments ($F=15.519, p<0.001$). Tukey HSD showed that \textit{DT3} grades were significantly lower than \textit{Touchalytics} and \textit{Breached!} ($p<0.001$), while \textit{Touchalytics} and \textit{Breached!} did not differ significantly. This lower performance is likely attributable to the exam administration mode. Unlike the other two courses, where students typed responses, \textit{DT3} DPs were administered on paper. This likely constrained response length and detail, a pattern consistent with prior work showing that typing often affords greater transcription fluency than handwriting~\cite{mogey2010handwriting, russell1999mode}. \textit{DT3} students wrote substantially less than Touchalytics and \textit{Breached!}. Because the grading rubric rewards explicit reasoning and justification, shorter responses were more likely to receive lower grades. Furthermore, word count showed a strong positive correlation with DP grades ($\rho=0.67, p<0.001$), suggesting that longer responses enabled students to better articulate and justify design choices. In contrast, time on task in the web-based courses showed a negligible correlation with grade ($\rho=-0.05, p=0.76$), suggesting that spending more time did not necessarily improve response quality.

\subsubsection{Student Perceptions of DPs}

As a validity check, inter-criterion correlations were negligible to moderate ($\rho=0.08$ to $0.46$), suggesting that students treated the five criteria as distinct dimensions rather than conflating them. We also found no evidence of a ``sour grapes'' effect: correlations between student grades and DP ratings were negligible to weak ($\rho=-0.06$ to $0.18$), indicating that student DP ratings were not driven by their performance.

Students' overall perceptions were positive, especially for HOT-related aspects of the DPs. \textit{Cognitive Complexity} received the highest mean rating ($M=4.35$), with 87\% agreeing that the prompts engaged HOT. \textit{Scenario Quality} was also rated highly ($M=4.21$), with 88\% finding the scenarios realistic and meaningful. In contrast, \textit{Feasibility} received the lowest rating ($M=3.63$), with 14\% expressing disagreement. However, this lower feasibility rating does not appear to reflect time pressure: average completion times in the web-based deployments were 10-14 minutes (well below the 30-minute limit), and the correlation between time on task and feasibility ratings was negligible ($\rho=0.11, p=0.498$), suggesting that students may have interpreted infeasibility as cognitive difficulty.

Students did not perceive significant differences between the standard and reasoning-oriented LLMs on any criterion (all $p>0.05$), unlike experts, who rated the reasoning-oriented LLM significantly higher on \textit{Scenario Quality} and \textit{Feasibility}. Comparing student and expert ratings on the 39 shared DPs, there was no agreement overall ($\kappa=-0.11$), indicating that the two groups often judged the same DPs differently. Students rated \textit{Scenario Quality} significantly higher than experts ($p=0.049$, $d=0.33$), but \textit{Feasibility} significantly lower ($p<0.001$, $d=0.74$). This feasibility gap is consistent with the ``expert blind spot'' hypothesis~\cite{nathan2003expert}, where experts may underestimate the cognitive load novices experience when solving unfamiliar, transfer-oriented tasks.

\subsubsection{Behavioral Patterns During DP-Solving}

Across the two web-based deployments (\textit{Breached!} and \textit{Touchalytics}), the mean \textit{initial planning latency} was about two minutes (167.2s for \textit{Breached!} and 114.6s for \textit{Touchalytics}). These values suggest that students spent time interpreting the scenario and organizing responses before typing. This pattern may suggest upfront \textit{analysis} of the DP, such as identifying constraints, clarifying requirements, and deciding how to structure responses, rather than immediately producing a memorized response.
The \textit{revision ratio} showed that students deleted 12-16 characters for every 100 characters typed, indicating that they interleaved writing and revision. This ongoing self-correction is consistent with evaluative thinking, as students checked whether their proposed designs fit the scenario, reconsidered wording or design choices, and refined justification while writing. Thus, revision behavior may suggest students were actively \textit{evaluating} emerging solutions rather than only transcribing already-decided answers.
Students also exhibited high \textit{burstiness} in both web-based exams, averaging 15 pauses per session, each longer than 10 seconds. While pauses can reflect multiple processes, their frequency may suggest repeated cognitive engagement during response construction, such as integrating design components, weighing tradeoffs, and \textit{synthesizing} a coherent solution.

\subsubsection{Relationship with Traditional Project Assessment}

Spearman’s $\rho$ between project and DP grades was negligible across all three courses (ranging from $-0.05$ to $0.03$). This lack of association suggests that DPs may capture dimensions of HOT in transfer contexts not reflected in traditional artifact-based assessment.
The learner-profile classification further clarifies this distinction. While most students (55.1\%) were \textit{Masters} who performed well in both assessments, a substantial group (28.2\%) were \textit{Implementers}: students with high project grades who scored below the passing threshold on the DP. Others were \textit{Conceptualizers} (10.3\%) or \textit{Strugglers} (6.4\%). This discrepancy suggests that strong project deliverables do not always translate to transfer-oriented HOT skills, and implementation success does not necessarily reflect individual mastery of design.
% \vspace{-1mm}

\section{Discussion and Limitations}
\label{section:discussion}

The negligible correlations between DP and project grades may suggest that DPs may capture transfer-oriented HOT not reflected in artifact-based assessment, supporting their role as complementary assessments. Behavioral patterns also suggest that students might be engaged in HOT while writing, aligning with the cognitive process theory of writing~\cite{flower1981cognitive,schilperoord2002pauses,TIAN2025100179}. The student-expert feasibility gap suggests that instructors should appropriately frame these open-ended tasks and balance their cognitive demands to reduce anxiety. The lower performance in the paper exam underscores the value of a digital delivery: handwriting may create a transcription barrier~\cite{mogey2010handwriting, russell1999mode} that constrains students' ability to fully demonstrate reasoning.

This study has several limitations. First, the expert evaluation was conducted by two authors, which may introduce confirmation bias. We mitigated this through rubric calibration, substantial inter-rater agreement, and triangulation with student ratings. Second, course comparisons are confounded by administration mode because \textit{DT3} used paper while the other courses used a web-based interface. Third, keystroke metrics are indirect behavioral proxies for planning, revision, and synthesis rather than direct measures of cognition. Finally, because DPs are open-ended, grading remains subjective and may require additional support to scale.
% \vspace{-2mm}

\section{Conclusion}
\label{section:conclusion}
This study proposes \textit{design problems} as a complementary assessment approach for PjBL, addressing a gap in the assessment of transfer-oriented HOT. Our findings show that LLMs, especially reasoning-oriented ones, can reliably generate realistic, LO-aligned, and cognitively demanding DPs across diverse projects. Expert evaluation and classroom deployment suggest that DPs can complement artifact-based assessment by eliciting individual reasoning in novel contexts, particularly when AI use and collaborative work may compromise student learning. By reducing the effort required to create project-specific DPs, LLMs can help scale this approach while leaving instructors responsible for review and adaptation.

% \vspace{-1mm}

\begin{acks}
This work is partially supported by the \grantsponsor{https://doi.org/10.13039/100000001}{U.S. National Science Foundation} - Award \grantnum{2515174}{DUE-2515174} and \grantnum{2513110} {DGE-2513110}.
\end{acks}

\bibliographystyle{ACM-Reference-Format} 

\balance
\bibliography{references}

\end{document}